# FPGA Implementation of a Reconfigurable Viterbi Decoder for WiMAX Receiver

Sherif Welsen Shaker
Nanoelectronics Integrated Systems Center, NISC, Nile University, Cairo, Egypt
swshaker@nileuniversity.edu.eg

Salwa Hussien Elramly
Ain Shams University
Cairo, Egypt
sramlye@netscape.net

Khaled Ali Shehata
Arab Academy for Science and Technology, AAST, Cairo, Egypt
k_shehata@aast.edu

*Abstract*— Field Programmable Gate Array technology (FPGA) is a highly configurable option for implementing many sophisticated signal processing tasks in Software Defined Radios (SDRs). Those types of radios are realized using highly configurable hardware platforms. Convolutional codes are used in every robust digital communication system and Viterbi algorithm is employed in wireless communications to decode the convolutional codes. Such decoders are complex and dissipate large amount of power. In this paper, a low power-reconfigurable Viterbi decoder for WiMAX receiver is described using a VHDL code for FPGA implementation. The proposed design is implemented on Xilinx Virtex-II Pro, XC2vpx30 FPGA using the FPGA Advantage Pro package provided by Mentor Graphics and ISE 10.1 by Xilinx.

## I. INTRODUCTION

SDR is characterized by its flexibility so that modifying or replacing software programs can completely change its functionality. SDRs can reduce the cost of manufacturing and testing, while providing a quick and easy way to upgrade the product and take the advantage of new signal processing techniques and new wireless phone applications [1,2]. In the early 1990's Field Programmable Gate Arrays (FPGAs) have become a considerable option in digital communication hardware where they were often applied as configurable logic cells to support memory controller tasks, complex state machines and bus interfacing [3]. Revolutionary changes have been made on FPGA technology in recent years. Complex real-time signal processing functions can yet be realized due to high clock speeds and huge gate densities provided by FPGA recent generations, like in Vertix-6 LXT FPGAs by Xilinx, which is optimized for high-performance logic and DSP with low power serial connectivity [4]. Many sophisticated signal processing tasks are performed in SDR that can be implemented on FPGA, including advanced compression algorithms, channel estimation, power control, forward error control, synchronization, equalization, and protocol management… etc [3].

Channel coding may be considered as one of the most challenging signal-processing tasks performed in SDR. Most digital communication systems use convoulutional coding to compensate for Additive White Gaussian Noise (AWGN) and the effects of other data degradation like channel fading and quantization noise. Viterbi decoding of convolutional codes has been found to be efficient and robust for the advantage that it has a fixed decoding time and it well suites to hardware decoding implementation [5]. The complexity of the Viterbi decoder grows exponentially with the increase of the number of the states of the convolutional encoder which relates to the constraint length of the corresponding encoder. Therefore, careful designs of the decoder have to be found to realize such a practical decoder. The aim of any decoder realization might be to reduce the area, reduce the power consumption or to enhance the performance. While reducing the area can be gained by the advances reached in electronics, low power designs have to be developed especially because they are important issues for mobile and portable applications. This paper proposes low power architecture for developing and modeling a Viterbi decoder for WiMAX software radio receiver. The design is described using a VHDL code for the implementation on FPGA hence it can be reconfigurable. It can also be fabricated as a low power Viterbi decoder on ASIC for either a base station or portable unit.

## II. VITERBI ALGORITHM

The Viterbi algorithm proposed by A.J. Viterbi is known as a maximum likelihood decoding algorithm for convolutional codes. So, it finds a branch in the code trellis most likely corresponds to the transmitted one. The algorithm is based on calculating the Hamming distance for every branch and the path that is most likely through the trellis will maximize that metric [5]. The algorithm reduces the complexity by eliminating the least likely path at each transmission stage. The path with the best metric is known as the survivor, while the other entering paths are non-survivors. If the best metric is shared by two or more paths, the survivor is selected from among the best paths at random.

The selection of survivors lies at the heart of the Viterbi algorithm and ensures that the algorithm terminates with the maximum likelihood path. The algorithm terminates when all of the nodes in the trellis have been labeled and their entering survivors are determined. We then go to the last node in the trellis and trace back through the trellis. At any given node, we can only continue backward on a path that survived upon entry into that node. Since each node has only one entering survivor, our trace-back operation always yields a unique path. This path is the maximum likelihood estimate that predicts the most likely transmitted sequence. [6].

$$P(Z/U^{(m')}) = \max P(Z/U^{(m)}) \text{ over all } U^{(m)} \quad (1)$$

where Z is the received sequence, and U(m) is one of the possible transmitted sequences, and chooses the maximum (closest possible received sequence).

Various coding schemes are used in wireless packet data network of different standards like GPRS, EDGE and WiMAX to maximize the channel capacity. GPRS uses a constraint length 5 and rate 1/2 Viterbi decoder, while EDGE uses a constraint length 7 and rate 1/3 with both tail biting Viterbi on the header portion and zero tail on data portion. WiMAX 802.16e, has the Viterbi with constraint length 7 and rate 1/2 with tail biting as mandatory and zero tail as optional [7]. This paper focuses on the implementation of the WiMAX Viterbi decoder with constraint length K = 7 and code rate r = 1/2 convolutional encoder as illustrated in figure 1. The trace-back depth depends mainly on the memory management of the algorithm. The longer the trace-back depth the larger the trellis will grow, and the larger the memory requirements. If the trace-back depth is also made too short, the performance of the codes will be affected significantly. An optimal trace-back depth of 5 * K may be used as stated in [6]. We consider the number of bits to be 40 per each frame. The tail sequence in each frame is fixed to 0's and resets the convolutional encoder to its initial state. Therefore, the trace-back operation can start from a known state.

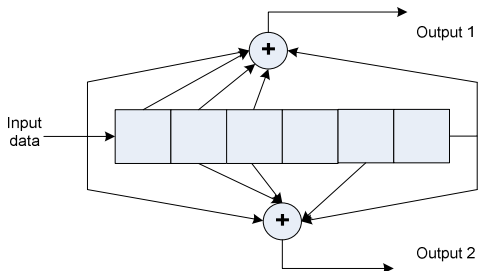

Fig. 1. K=7, r=1/2 convolutional encoder

### III. ARCHITECTURE OF VITERBI DECODER

The input to the communication receiver is a continuous stream of analog, modulated signals. The primary tasks of the receiver are the recovery of the carrier and bit timing so that the individual received data bits can be removed from the carrier and separated from one another in an efficient manner. Both tasks are generally performed through the use of phase-locked loops. The analog baseband signals are then applied to the analog-to-digital converter with b-bit quantizer to get a received bit stream. Then the bit stream is applied as the input to the Viterbi decoder. In order to compute the branch metrics at any given point in time, the Viterbi decoder must be able to segment the received bit stream into n-bit blocks, each block corresponding to a stage in the trellis. In this paper, we assume that the input to our proposed design is an identified code symbols and frames, i.e. the design decodes successive bit stream. The Viterbi decoder consists of some basic building blocks as illustrated in figure 2.

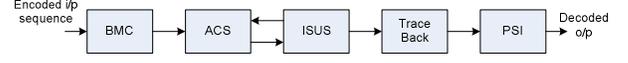

Fig. 2. Basic building blocks for Viterbi decoder

#### A. The Branch Metric Computer (BMC)

This is typically based on a look-up table containing the various bit metrics. The computer looks up the n-bit metrics associated with each branch and sums them to obtain the branch metric. The result is passed along to the path metric update and storage unit. Figure 3 shows the block diagram of the BMC [8].

#### B. The Path Metric Updating and Storage

This takes the branch metrics computed by the BMC and computes the partial path metrics at each node in the trellis. The surviving path at each node is identified, and the information-sequence updating and storage unit notified accordingly. Since the entire trellis is multiple images of the same simple element, a single circuit called Add-Compare-Select may be assigned to each trellis state.

#### C. Add-Compare-Select (ACS)

ACS is being used repeatedly in the decoder [6]. The entire decoder can be based on one such ACS, resulting in a very slow, but low-cost, implementation. On the other hand, a separate ACS circuit can be dedicated to every element in the trellis, resulting in a fast, massively parallel implementation. For a given code with rate 1/n and total memory M, the number of ACS required to decode a received sequence of length L is $L \times 2^M$. In our implementation we combined both the BMC and the ACS in one unit representing a single wing of each trellis butterfly as illustrated in figure 3.

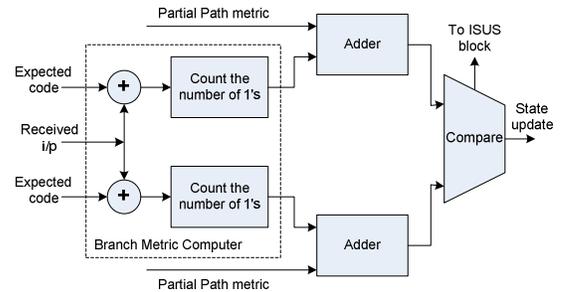

Fig. 3. ACS module

#### D. The Information Sequence Updating and Storage (ISUS)

This is responsible for keeping track of the information bits associated with the surviving paths designated by the path metric updating and storage unit. There are two basic design approaches: Register Exchange and Trace Back. In both techniques, a shift register is associated with every trellis node throughout the decoding operation [9]. Each state is assigned a register which has a length equal to the frame length. The register exchange may offer high-speed operation but it is not power efficient while the trace-back is complex but has the advantage of low power dissipation because the switching activity of the registers is much lower than that of the register exchange. Since one of the major interests is the low power design, the proposed decoder in this paper has been implemented using the trace back approach.

Our proposed ISUS is shown in figure 4, where for the encoder with constraint length K=7 and code rate 1/2, 40 registers are needed, each of them with size 64 bits. The survivor branch information is computed as following:

For the current state, if the input symbol is coming from the upper branch then set the corresponding state flip-flop to 1 in the 64-bit register. If the input symbol is coming from the lower branch then clear the state flip-flop. The bit in the corresponding flip-flop for the current state is being used associated with the current state itself to determine what the previous state was.

As time progresses, the survivor branch information is filled into the registers from left to right. The key issue is that the content of each register does not change as soon as it is updated. This is a very useful in our low power design, as we don't have to activate the registers after updating hence the switching activity reduces and of course a reduction in power dissipation can be reached. This can be realized by the clock gating-scheme [10]. As shown in figure 4, the clock of each register is enabled only when the register updates its survivor path information, while the clock of all other registers is disabled ensuring them to hold their states. The clock is gated by the information coming from a ring counter that tells what the current state is so far. The current state of the counter is as the same as the number of the received code symbols.

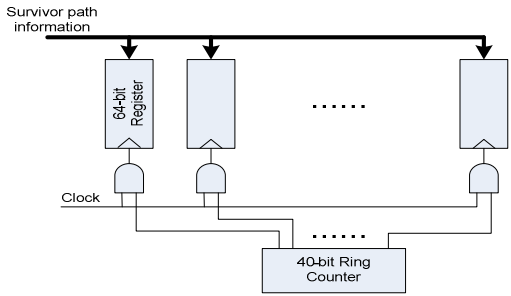

Fig. 4. Information sequence Update and Storage block diagram

*E. The Trace Back unit*

This is responsible for tracing back the survivor path after receiving the entire frame. The tracing back operation uses the bit in the corresponding flip-flop of the current state with the current state itself to determine the previous state; this is illustrated as in figure 5 and can be denoted as the following:

*For the current state $S_{2j}$, if the state input is from upper branch, then the previous state was $S_{j+32}$. Else, the previous state was $S_j$. For the current state $S_{2j+1}$, if the state input is from upper branch, then the previous state was $S_{j+32}$. Else, the previous state was $S_j$*

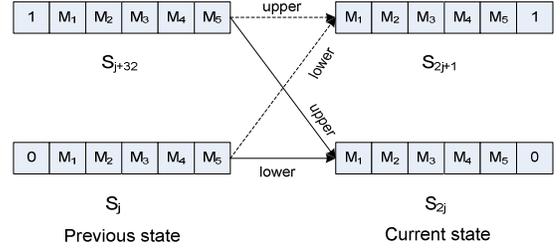

Fig. 5. Trace back unit decision rule

After tracing back the survivor path at the end of each frame, the decoded output sequence is being generated as the following:

*For odd states $S_1$, $S_3$ …, $S_{63}$ decide the input to the current state is symbol 1. For even states $S_0$, $S_2$ …, $S_{62}$ decide the input to the current state is symbol 0.* Since the decoding circuits have no need to work until all the frame symbols is being received, In our proposed design we enable those circuits one clock at the end of each frame, this saves more power. The survivor is then applied to a Parallel-to-Serial Interface that is implemented by 40-bit parallel-to-serial shift register.

## IV. RESULTS

A Matlab code is first driven to evaluate the performance for the K=7, rate 1/2 Viterbi decoder. Figure 6 shows the BER curve vs. $E_b/N_o$ using AWGN channel for both the uncoded BPSK and the convolutional one with Viterbi decoding. Both the convolutional encoder with constraint length 7 and code rate 1/2 and our proposed Viterbi decoder are then described at the register transfer level by VHDL code, that consider the low power design technique using the concept of trace-back approach along with clock gating and. A VHDL code is also written for the Viterbi decoder with trace-back and shift update approach to be compared in the power dissipation with the proposed design. The test-bench is written for both the encoder and the decoder as unique system, to provide the clock, control signals and the data to be encoded and then decoded. Noise generator module is implemented for the purpose of simulating the decoder; it simply adds noise by inverting some bits in transmitted frames. The test bench block diagram for the proposed decoder is shown in figure 7.

The simulation of the proposed Viterbi decoder has been made using Modelsim SE 6.4b digital simulator is then made to test the function of the implemented decoder. Figure 8 shows the simulation waveforms with applying an error pattern of 7 bit in error along the received frame. The result shows the efficient performance of the decoder in correcting the 7 errors. As it can be seen, the decoded sequence will not be available until the end of the frame. After receiving the first frame, the data is kept decoded every clock cycle but delayed 40 clocks. This feature is very helpful in our design, since the trace back module does not work until the end of each frame and hence we don't have to activate the trace back module until the end of the frame and only for one clock cycle, by this feature, more saving in dissipated power.

Fig. 6. Matlab simulation for K=7, r=1/2 Viterbi decoder

Fig. 7. Viterbi decoder, test bench

Fig. 8. Simulation of the decoder, 7-bit error pattern is corrected

Xilinx ISE 10.1 tools has been used for the synthesization purposes to map the design to the FPGA target technology. Xilinx Virtex-II Pro, xc2vpx30, with speed grade -6 has been selected; the design took about less than (7%) of the total chip resources. The results also showed that the proposed design can work with frequency up to 47. 4 MHz. The proposed design and the one written using the shift update in trace-back module are then applied to Xilinx XPower analyzer tool. Table I provides the power dissipation for both designs at clock frequency 12 MHZ and supply voltage Vccint=1.5V. The results in table I shows how the dynamic power is being minimized for the proposed design using the trace-back approach with clock gating and disabling the decoding circuits compared with other design that uses the shift update and does not benefits the gating. The results indicates the power reduction also when shifting the design on ASIC for using in base station.

TABLE I. COMPARISON OF POWRE DISSIPATION

|  | *Design using Trace back with shift update* | *Proposed Design* |
|---|---|---|
| ***Total Quiescent Powre*** | 0.181 W | 0.054 W |
| ***Total Dynamic Power*** | 0.050 W | 0.012 W |
| ***Total Power*** | 0.231 W | 0.065 W |

V. CONCLUSION

Features like higher flexibility, re-configurability and shorter time-to-market give the FPGA new opportunities for the effective insertion in SDR conditioning chain. In this paper, a design of low-power, configurable Viterbi decoder for WiMAX receiver has been proposed. The design benefits the concept of trace-back approach with clock gating for power reduction. The design has been described by VHDL using FPGA Advantage Pro by Mentor Graphics, simulated using Modelsim SE 6.4b, and then targeted on Xilinx Virtex-II Pro, XC2vp30 FPGA using ISE design suit 10.1 by Xilinx. The power report for the proposed design have been driven using Xilinx XPower Analyzer tool, and it showed the reduction in the dynamic power dissipation which is a good indication for power reduction when implementing the proposed design on ASIC. The design took about 7% of the total chip logic elements. The maximum operating frequency is 47.4 MHz, which is found adequate to our application.